\documentclass{article}%
\usepackage{amssymb}
\usepackage{amsfonts}%
\usepackage{amsmath}%
\usepackage{graphicx}
\providecommand{\U}[1]{\protect\rule{.1in}{.1in}}

\begin{document}

\title{Saturation in Deep Inelastic Scattering \\from AdS/CFT}
\author{Lorenzo Cornalba$^{a}$, Miguel S. Costa$^{b}$ \medskip\\$^{a}$ Centro Studi e Ricerche E. Fermi\\Compendio Viminale, I-00184, Roma\\Universit\`{a} di Milano-Bicocca and INFN, sezione di Milano-Bicocca\\Piazza della Scienza 3, I-20126 Milano, Italy \medskip\\$^{b}$ Departamento de F\'{\i}sica e Centro de F\'{\i}sica do Porto,\\Faculdade de Ci\^{e}ncias da Universidade do Porto,\\Rua do Campo Alegre 687, 4169-007 Porto, Portugal \medskip\\{\small {\texttt{Lorenzo.Cornalba@mib.infn.it, miguelc@fc.up.pt}}}}
\date{}
\maketitle

\begin{abstract}
We analyze deep inelastic scattering at small Bjorken $x$, using the
approximate conformal invariance of QCD at high energies. Hard pomeron
exchanges are resummed eikonally, restoring unitarity at large values of the
phase shift in the dual AdS geometry. At weak coupling this phase is
imaginary, corresponding to a black disk in AdS. In this saturated regime,
cross sections exhibit geometric scaling and have a simple universal form,
which we test against available experimental data for the proton structure
function $F_{2}\big( x,Q^{2}\big)$. We predict, in particular, the dependence
of the cross section on the scaling variable $(Q/Q_{\mathrm{s}})^{2}$ in the
deeply saturated region, where $Q_{\mathrm{s}}$ is the usual saturation
scale. We find agreement with current data on $F_{2}$ in the 
kinematical region $0.5<Q^{2}<10$ GeV$^{2}$, $x< 10^{-2}$, with an average
$6\%$ accuracy. We conclude by discussing the relation of our approach with
the commonly used dipole formalism.

\end{abstract}

\section{Introduction}

The high energy behavior of QCD is greatly simplified by the asymptotic
weakness of the coupling and the approximate conformal invariance of the
theory. Of great interest, in this respect, is the study of interaction
processes in the Regge limit of high center of mass energy, with the other
kinematical invariants kept fixed. This kinematical regime is, for instance,
relevant to the analysis of deep inelastic scattering (DIS) experiments at
fixed photon virtuality $Q^{2}$ in the limit of vanishing Bjorken $x$.

\begin{figure}
\begin{center}
\includegraphics[height=5cm]{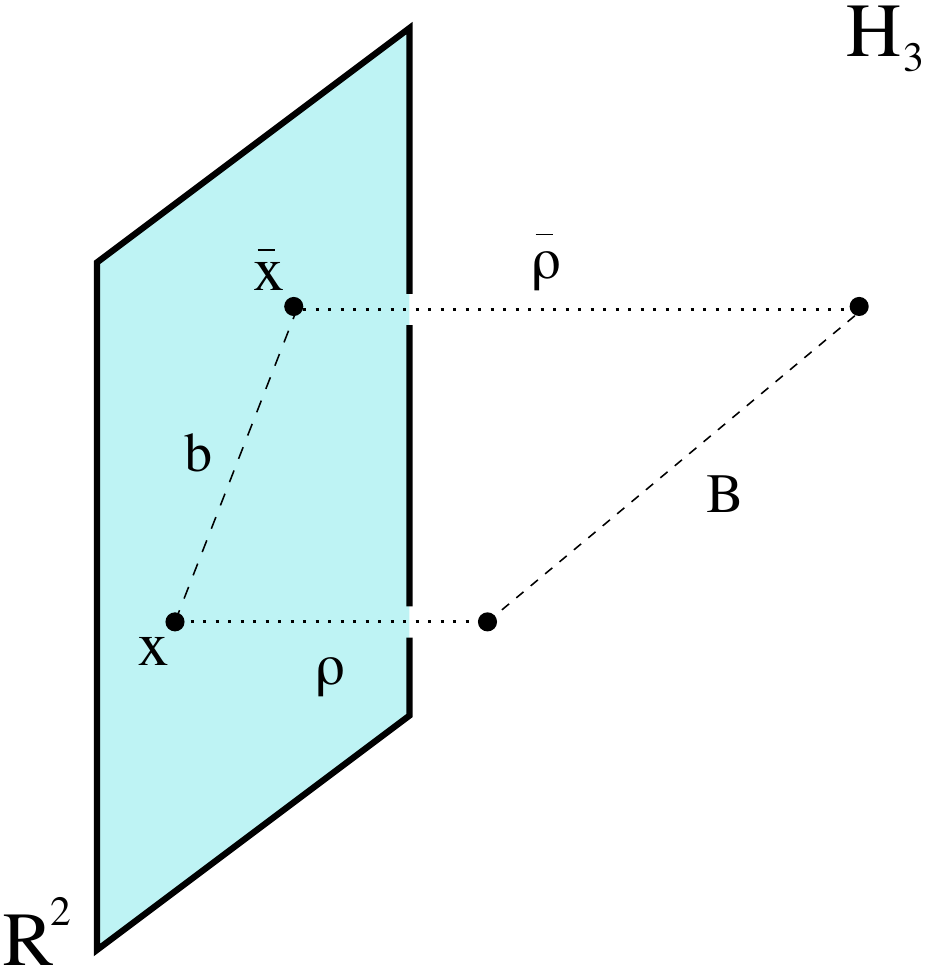}
\caption{
Impact points $\rho,\mathbf{x}$ and $\bar{\rho},\mathbf{\bar{x}}$ in \textrm{H}$_{3}$, 
separated by a geodesic distance $B$. 
Dropping the radial coordinates $\rho$ and $\bar{\rho}$, we obtain the impact points
$\mathbf{x}$ and $\mathbf{\bar{x}}$ in the  plane transverse to the high energy process \textrm{R}$^2$ , 
with impact parameter $\mathbf{b}=\mathbf{x}-\mathbf{\bar{x}}$.}
\end{center}
\label{holography}
\end{figure}

To the extent that QCD can be approximated by a conformal field theory, we
must in general analyze CFT correlators of the form
\begin{equation}
\left\langle \mathcal{O}_{1}(q_{1})\, \mathcal{O}_{2}(q_{2})\, \mathcal{O}%
_{1}^{\star}(q_{3}) \,\mathcal{O}_{2}^{\star}(q_{4}) \right\rangle
\label{eq100}%
\end{equation}
in the limit of large $s=-\left(  q_{1}+q_{2}\right)  ^{2}$, and at fixed
virtualities $Q_{i}^{2}=q_{i}^{2}$ and momentum transfer $t=-\left(
q_{1}+q_{3}\right)  ^{2}$. In the high energy limit, the correlator
(\ref{eq100}) is best analyzed in impact parameter space. The correct
representation is suggested by the AdS/CFT duality \cite{AdSCFT}, 
although  let us stress that all of the results in this paper
are purely based on simple implications of conformal symmetry and could be
derived also in the field theory language. Considering 
(\ref{eq100}) as a high energy process in AdS$_{5}$, the relevant
transverse space is then the three dimensional hyperbolic space $\mathrm{H}%
_{3}$, holographically dual to the usual two--dimensional plane transverse to
the high energy process described by (\ref{eq100}), as shown in Figure
\ref{holography}. Representing four dimensionl vectors as $\left(  x^{+}%
,x^{-},\mathbf{x}\right)  $, with $x^{\pm}$ lightcone variables and
$\mathbf{x}$ a two dimensional transverse vector, we can parameterize
$\mathrm{H}_{3}$ using Poincar\'{e} coordinates $\rho,\mathbf{x}$ with metric
$\rho^{-2}\left(  d\rho^{2}+d\mathbf{x}^{2}\right)  $ and volume form
$\rho^{-3}d\rho~d^{2}\mathbf{x}$, with $\rho>0$ the distance to the
holographic boundary of $\mathrm{H}_{3}$. Following the results in \cite{PS0,
PS1,Paper1,Paper2,Paper3,PS2,PS3,Paper4,Paper5}, we may write the impact
parameter representation for the correlator (\ref{eq100}). Choosing, for
simplicity of exposition, external scalar operators, it is given by
\begin{equation}
2s\int d^{2}\mathbf{b~}e^{i\mathbf{b\cdot q}}~e^{2i\delta( s,\mathbf{b})
}~,\label{eq200}%
\end{equation}
where $\mathbf{q}$ is the transverse momentum transfer with $-t=\mathbf{q}%
^{2}$ and where $\mathbf{b}$ is the usual impact parameter. The phase shift
$\delta(s,\mathbf{b}) $ is itself given by
\begin{equation}
e^{2i\delta( s,\mathbf{b}) }= \int\frac{d\rho}{\rho^{3}}\, f_{1}( \rho)\,
f_{3}( \rho) \int\frac{d\bar{\rho}}{\bar{\rho}^{3}}\, f_{2}( \bar{\rho})\,
f_{4}( \bar{\rho})\; e^{2i\Delta( S,B) }\ ,\label{eq300}%
\end{equation}
with $\Delta( S,B) $ the phase shift in AdS, which depends on the AdS energy
squared and impact parameter $S$ and $B$, according to\footnote{We take the
AdS quantities $S$ and $B$ to be dimensionless, measured in units of the AdS
radius.}
\begin{align}
S  & =\rho\bar{\rho}s~,\nonumber\\
\cosh B  & =\frac{\rho^{2}+\bar{\rho}^{2}+\mathbf{b}^{2}}{2\rho\bar{\rho}%
}~.\label{eq1000}%
\end{align}
In particular, $B$ is the geodesic distance between the points $\rho
,\mathbf{x}$ and $\bar{\rho},\mathbf{\bar{x}}$ in \textrm{H}$_{3}$, with
$\mathbf{b}=\mathbf{x}-\mathbf{\bar{x}}$. These represent the impact points of
the operators $\mathcal{O}_{1}$ and $\mathcal{O}_{2}$ in the transverse space.
Finally, the functions $f_{i}$ are the radial wave functions for the
scattering states. For scalar operators $\mathcal{O}_{1},\mathcal{O}_{2}$ of
dimension $\Delta_{1},\Delta_{2}$ they are given by $f_{i}\propto Q_{i}%
\rho^{2}K_{\Delta_{1}-2}( Q_{i}\rho)$ for $i=1,3$, and by $f_{i}\propto
Q_{i}\bar{\rho}^{2}K_{\Delta_{2}-2}( Q_{i}\bar{\rho})$ for $i=2,4$
\cite{AdSCFT2}. We normalize the wavefunctions so that
\begin{equation}
\int\frac{d\rho}{\rho^{3}}\, f_{1}( \rho) \,f_{3}( \rho) =\int\frac{d\bar
{\rho}}{\bar{\rho}^{3}}\, f_{2}( \bar{\rho})\, f_{4}( \bar{\rho})
=1~.\label{eq7000}%
\end{equation}

As shown in \cite{Paper2}, the impact parameter representation (\ref{eq200})
and (\ref{eq300}) approximates the conformal partial wave decomposition of the
correlator (\ref{eq100}) in the channel $\mathcal{O}_{1}\mathcal{O}%
_{2}\rightarrow\mathcal{O}_{1}^{\star}\mathcal{O}_{2}^{\star}\,$, with
intermediate states of conformal dimension and spin respectively given by
$\sqrt{S}\cosh( B/2) $ and $\sqrt{S}\sinh( B/2) $. In analogy with the usual
results for scattering in flat space, we then expect that AdS unitarity
implies \cite{Paper4, Paper5}
\[
\operatorname{Im}\Delta( S,B) \geq0~,
\]
even though the phase shift $\delta( s,\mathbf{b}) $ does not satisfy a simple
unitarity constraint.

We shall focus, for concreteness, on the very relevant and simple case of
vanishing momentum transfer $\mathbf{q}=0$ and equal virtualities for the
incoming and outgoing states $Q=Q_{1}=Q_{3}$ and $\bar{Q}=Q_{2}=Q_{4}$. It is
then natural to construct, from the correlator (\ref{eq200}), the following
effective cross section%
\[
\Sigma\big( s,Q,\bar{Q}\big) = 2\int d^{2}\mathbf{b~}\operatorname{Re} \left(
1-e^{2i\delta( s,\mathbf{b}) }\right)  ~.
\]
Using (\ref{eq300}), the cross section $\Sigma$ can be conveniently written as%
\begin{equation}
2\int\frac{d\rho}{\rho^{3}}\, f_{1}( \rho) f_{3}( \rho) \int\frac{d\bar{\rho}%
}{\bar{\rho}^{3}}\, f_{2}( \bar{\rho})\, f_{4}( \bar{\rho}) \; \sigma(
s,\rho,\bar{\rho}) ~,\label{eq600}%
\end{equation}
where we have defined the unintegrated cross sections%
\begin{align}
\sigma( s,\rho,\bar{\rho})  & = \int d^{2}\mathbf{b~}\sigma( s,\rho,\bar{\rho
},\mathbf{b}) ~,\label{eq700}\\
\sigma( s,\rho,\bar{\rho},\mathbf{b})  & = \operatorname{Re}\left(
1-e^{2i\Delta( S,B) }\right)  ~.\nonumber
\end{align}
In this language, $\sigma( s,\rho,\bar{\rho},\mathbf{b}) $ is the natural
object which automatically satisfies the unitarity bound $0\leq\sigma\leq2$
due to AdS unitarity $\operatorname{Im}\Delta\geq0$. Moreover, for a black
disk region we have $\sigma\rightarrow1$, corresponding to a phase shift
$\Delta$ with large imaginary part.

In general, we cannot evaluate the integral over the impact parameter
$\mathbf{b}$. We may, on the other hand, use the\ relation (\ref{eq1000})
between $\mathbf{b}$ and the AdS impact parameter $B$ to rewrite the cross
section $\sigma( s,\rho,\bar{\rho}) $ in (\ref{eq700}) as%
\begin{equation}
\sigma( s,\rho,\bar{\rho}) = 2\pi\rho\bar{\rho} \int_{\left\vert \ln(
\bar{\rho}/\rho) \right\vert }^{\infty}dB~\sinh B~\operatorname{Re}\left(
1-e^{2i\Delta( s\rho\bar{\rho},B)}\right)  ~.\label{eq1100}%
\end{equation}
It is now apparent that we are probing the phase $\Delta( S,B) $ for fixed
$S=s\rho\bar{\rho}$ and for $B\geq\left\vert \ln(\bar{\rho}/\rho)\right\vert
$. Finally, note that the unintegrated cross sections satisfy, due to
conformal invariance, non trivial relations under the transformation
$\rho\rightarrow\bar{\rho}^{2}/\rho$ with $s\rightarrow s\left(  \rho^{2}%
/\bar{\rho}^{2}\right)  $, $\mathbf{b\rightarrow b}\left(  \bar{\rho}%
/\rho\right)  $, which leave invariant $S$ and $B$. More precisely, $\sigma(
s,\rho,\bar{\rho},\mathbf{b})$ is invariant whereas
\[
\frac{\rho^{2}}{\bar{\rho}^{2}}\, \sigma\left(  s\,\frac{\rho^{2}}{\bar{\rho
}^{2}},\frac{\bar{\rho}^{2}}{\rho},\bar{\rho}\right)  =\sigma( s,\rho
,\bar{\rho}) ~.
\]

The phase shift $\Delta( S,B) $ depends both on the number of colors $N$ and
on the 't Hooft coupling $\bar{\alpha}_{s}=\alpha_{s}N/\pi$ of the theory. For
large energy squared and impact parameter $S$ and $B$, the phase $\Delta$ will
be dominated by the leading Regge pole of the planar diagrams of the theory
\cite{Paper4, Paper5}, and will have a general representation of the form\footnote{In this paper, 
in order to have a uniform notation, we use slightly
different conventions then in \cite{Paper4, Paper5}. In particular,
$\Delta=-\pi\Gamma_{\text{there}}$ and $\beta_{\text{here}}=-\pi
\beta_{\text{there}}$.}%
\begin{equation}
\Delta( S,B) =\frac{1}{N^{2}}\int d\nu~\beta( \nu) ~S^{j( \nu) -1}%
~\Omega_{i\nu}( B) ~,\label{eq400}%
\end{equation}
where the Regge spin $j( \nu) $ and residue $\beta( \nu) $ depend implicitly
only on the 't Hooft coupling $\bar{\alpha}_{s}~$and are even functions of
$\nu$. The function $\Omega_{i\nu}(B)$ computes radial Fourier transforms in
$\mathrm{H}_{3}$, satisfies $\left(  \square_{\mathrm{H}_{3}}+\nu
^{2}+1\right)  \Omega_{i\nu}=0$ and is given explicitly by
\begin{equation}
\Omega_{i\nu}( B) =\frac{1}{4\pi^{2}}\frac{\nu\sin\nu B}{\sinh B}~.\nonumber
\end{equation}

Whenever the AdS phase satisfies $\left\vert \Delta\right\vert \ll1$, the full
cross section is well approximated by a single Reggeon exchange, and we may
write
\begin{equation}
\sigma( s,\rho,\bar{\rho},\mathbf{b}) \simeq2\operatorname{Im}\Delta( S,B)
.\label{eq4000}%
\end{equation}
In this case, the integral over the impact parameter $\mathbf{b}$ can be
explicitly performed. In fact, using the Regge representation (\ref{eq400})
for the phase shift, together with\footnote{In order to correctly compute the
normalization of this Fourier transform, as well as of the ones in the sequel
of the paper, it is safest to compute at non--zero momentum transfer and take
the limit $\mathbf{q\rightarrow0}$.}%
\[
\int d^{2}\mathbf{b}~\Omega_{i\nu}( B) =\frac{1}{2\pi}\,\rho\bar{\rho}\left(
\frac{\bar{\rho}}{\rho}\right) ^{-i\nu}~,
\]
coming from the integral representation \cite{Paper5}%
\[
\Omega_{i\nu}( B) = \frac{\nu^{2}}{4\pi^{3}}\int d^{2}\mathbf{z} \left(
\frac{\rho}{\rho^{2}+\left(  \mathbf{b}-\mathbf{z}\right)  ^{2}} \right)
^{1+i\nu}\left(  \frac{\bar{\rho}}{\bar{\rho}^{2}+\mathbf{z}^{2}} \right)
^{1-i\nu}~,
\]
we may evaluate the cross section $\sigma( s,\rho,\bar{\rho}) $ to be
\begin{equation}
\sigma( s,\rho,\bar{\rho}) \simeq\frac{\rho\bar{\rho}}{~2\pi\,N^{2}%
}\,\operatorname{Im} \int d\nu~\beta( \nu) \; ( s\rho\bar{\rho})^{j( \nu)
-1}\, \left(  \frac{\bar{\rho}}{\rho}\right)  ^{-i\nu}~.\label{eq7200}%
\end{equation}

\section{The Cross Section Deep into Saturation\label{secDeep}}

At fixed AdS energy squared $S$, the phase $\Delta( S,B) $ will in general
vanish in the limit $B\rightarrow\infty$. On the other hand, as we approach
smaller and smaller impact parameters, $\Delta$ will in general grow and reach
saturation at $B\simeq\mathrm{B}_{\mathrm{s}}( S) $, where $\Delta$ is of
order one.

\begin{figure}
\begin{center}
\includegraphics[height=5cm]{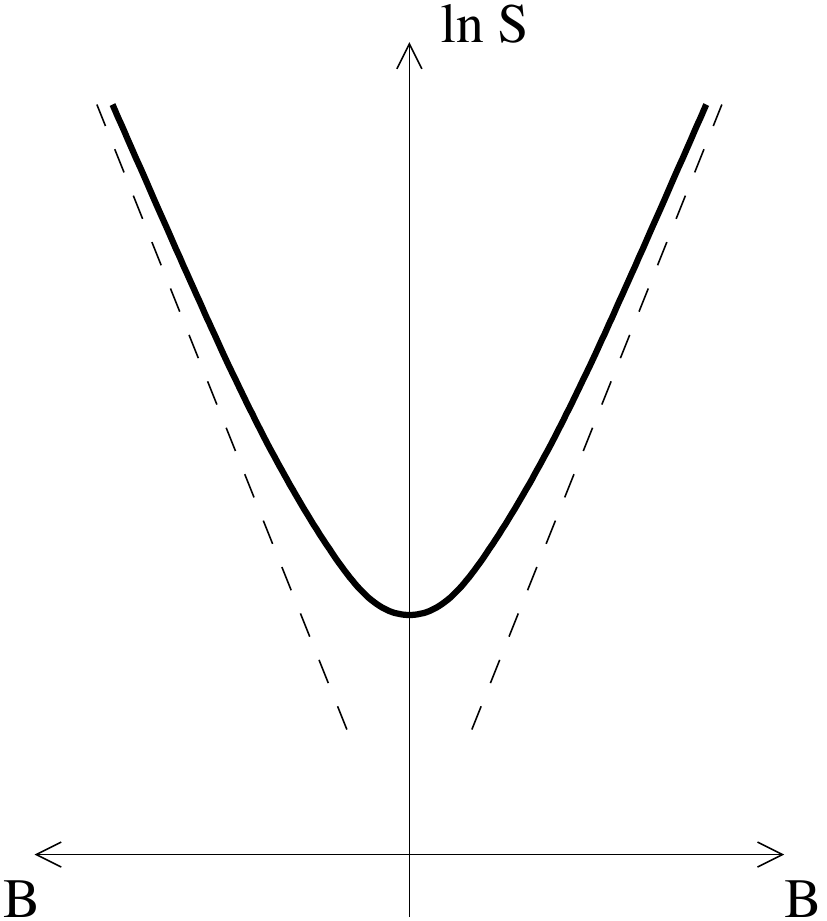}
\end{center}
\caption{Saturation line B$_s(S)$ in the $B$--$\ln(S)$ plane. As we increase $S$, the saturation
line starts at a minimal value of $S$ of the order of $\bar{\alpha}_s^{-1}$ and reaches the asymptotic linear behavior,
shown with a dashed line, for large $S$. We extended the graph to the left of the $B=0$ axis symmetrically, drawing the mirror image of the saturation line. This is convenient since $B=B(\rho,\bar{\rho},b)$ is invariant under $\rho\leftrightarrow \bar{\rho}$ and we wish to show
separately regions with $\rho>\bar{\rho}$ and $\rho<\bar{\rho}$. }
\label{lnS-B}
\end{figure}

We will be mostly interested in weakly coupled gauge theories, where the phase
is predominantely imaginary \cite{BFKL}.\footnote{On the other hand, at strong coupling and for large
impact parameters the phase shift is predominantly real and is given by the
gravi-reggeon exchange in AdS. Studies of DIS in this regime include \cite{DISstrongcoupling}.
Saturation effects at strong coupling have also been analyzed in  \cite{Saturationstrongcoupling}, 
and a conjectured relation to black hole formation was put forward in \cite{BHcollapse}.} In this case, saturation is reached
at\footnote{Note that it is usually believed that, when $\operatorname{Im}\Delta\simeq1$, non--linear BK corrections
to $\Delta$ of order $N^{-4}$ due to fan diagrams also become relevant \cite{MuellerLectures}. As long as those corrections are negligible for impact parameters
larger then the saturation line, they are irrelevant in the discussion which follows, since they will predominantly affect the phase shift 
in the black disk region.}
\[
2\operatorname{Im}\Delta\big( S,\mathrm{B}_{\mathrm{s}}( S) \big)\simeq1~.
\]
A typical plot of the saturation line in the $\big( B,\ln S\big) $ plane is
given in Figure \ref{lnS-B}. In particular, for large $S$ we have the 
linear relation
\begin{equation}
\mathrm{B}_{\mathrm{s}}( S) \simeq\omega~\ln S+\cdots~,\label{eq1200}%
\end{equation}
where $\cdots$ represents subleading terms in $S$. This can be shown, as
customary \cite{satLine}, by approximating the integral in (\ref{eq400}) at the saddle point
$iB=j^{\prime}( \nu) \ln S$. Saturation is then reached when the phase in
(\ref{eq400}) vanishes at the saddle -- i.e. when $\left( 1+i\nu_{\mathrm{s}%
}\right)  \mathrm{B}_{\mathrm{s}}= \left(  j( \nu_{\mathrm{s}}) -1\right)  \ln
S$. These conditions imply that
\[
\omega=-i\;j^{\prime}( \nu_{\mathrm{s}})
\]
where the saturation saddle point $\nu_{\mathrm{s}}$ is defined in terms of
the Regge trajectory $j( \nu) $ by
\[
\left(  1+i\nu_{\mathrm{s}}\right)  \,\omega=j( \nu_{\mathrm{s}}) -1~.
\]

The cross section $\sigma( s,\rho,\bar{\rho}) $ near saturation $\left\vert
\ln( \bar{\rho}/\rho) \right\vert \gtrsim\mathrm{B}_{\mathrm{s}}( s\rho
\bar{\rho})$ exhibits geometric scaling \cite{geometric}. More precisely, the
integral (\ref{eq7200}) has a leading behavior given by
\begin{equation}
\sigma( s,\rho,\bar{\rho}) \sim\bar{\rho}^{2}~\tau^{-\left(  1+i\nu
_{\mathrm{s}}\right)  \frac{\left(  1-\omega\right)  }{2}}~,\label{eq8000}%
\end{equation}
where we have defined the scaling variable%
\begin{equation}
\tau=\frac{\bar{\rho}^{2}}{\rho^{2}}\left(  s\rho^{2}\right)  ^{-\frac
{2\omega}{1-\omega}}~.\label{eq8100}%
\end{equation}
On the other hand, we are interested in the analysis of the cross section
$\sigma( s,\rho,\bar{\rho})$ deep inside saturation, that is for
\begin{equation}
\left\vert \ln( \bar{\rho}/\rho) \right\vert \lesssim\mathrm{B}_{\mathrm{s}}(
s\rho\bar{\rho}) ~.\label{eq5000}%
\end{equation}
In this case, the integral (\ref{eq1100}) is dominated by the region
$B\lesssim\mathrm{B}_{\mathrm{s}}$, where we may replace $\sigma(s,\rho
,\bar{\rho},\mathbf{b}) \simeq1$. This situation corresponds to a simple black
disk in AdS,\ even though this is less transparent from the four dimensional
perspective. We then obtain the approximate expression for the cross section
\begin{align}
\sigma( s,\rho,\bar{\rho})  & \simeq 2\pi\rho\bar{\rho} \int_{\left\vert \ln(
\bar{\rho}/\rho) \right\vert }^{\mathrm{B}_{\mathrm{s}}}dB~\sinh B\nonumber\\
& \simeq\pi\rho\bar{\rho}\,\left[  2\cosh\mathrm{B}_{\mathrm{s}}(s\rho
\bar{\rho}) -\frac{\rho}{\bar{\rho}}-\frac{\bar{\rho}}{\rho}\right]
~.\label{eq5100}%
\end{align}
Moreover, when $S=s\rho\bar{\rho}$ is large and we are in the linear regime
(\ref{eq1200}), we have the simpler approximate expression
\begin{equation}
\sigma( s,\rho,\bar{\rho}) \simeq~\pi\rho\bar{\rho}\, \left[ \left(  s\rho
\bar{\rho}\right)  ^{\omega} +\left(  s\rho\bar{\rho}\right) ^{-\omega}%
-\frac{\rho}{\bar{\rho}}-\frac{\bar{\rho}}{\rho} \right]  ~,\label{eq2000}%
\end{equation}
where we neglect any subleading term in (\ref{eq1200}). Note that, for
$\mathrm{B}_{\mathrm{s}}\gg\left\vert \ln( \bar{\rho}/\rho) \right\vert \gg1$
equation (\ref{eq2000}) is dominated by the first term and we obtain
\[
\sigma( s,\rho,\bar{\rho}) \sim\bar{\rho}^{2}~\tau^{\,-\frac{1-\omega}{2}}~,
\]
to be contrasted with (\ref{eq8000}), valid near saturation. Finally note that, even 
in the deeply saturated region, the cross section $\sigma( s,\rho,\bar{\rho})$ grows
with $s$ with a power law, violating the Froissart bound. Recall though that (\ref{eq2000})
has been derived by assuming an exact conformal symmetry and that, in a conformal theory,
the Froissart bound is not relevant since there is no mass scale.

For the sake of clarity, let us discuss a specific example, which is at the
same time simple and instructive, since it contains most of the relevant
physics. We will work with the maximally superconformal version of QCD,
$\mathcal{N}=4$ SYM with $SU\left(  N\right)  $ gauge group, and we will
consider the scalar operators $\mathcal{O}_{1}=\mathrm{Tr}\left(
Z^{2}\right)  $ and $\mathcal{O}_{2}=\mathrm{Tr}\left(  W^{2}\right)  $ of
dimension $\Delta_{1}=\Delta_{2}=2$, with $Z$ and $W$ two of the three complex
adjoint scalars of the theory. To leading order in $\bar{\alpha}_{s}$, we have
the well known BFKL result \cite{BFKL}%
\[
j( \nu) \simeq1+\bar{\alpha}_{s}\left(  2\Psi(1)-\Psi\left(  \frac{1+i\nu}%
{2}\right)  -\Psi\left(  \frac{1-i\nu}{2}\right)  \right)
\]
and \cite{Paper5}
\begin{equation}
\beta(\nu) \simeq i~16\pi^{4}\bar{\alpha}_{s}^{2}~ \frac{\tanh\frac{\pi\nu}%
{2}}{\nu\left(  1+\nu^{2}\right)  ^{2}}~.\label{eq10}%
\end{equation}
At vanishing $\ln S$, the integral (\ref{eq400}) can be explicitly computed to
be
\begin{align*}
\Delta( S=1,B)  & = \frac{i}{3}\,\alpha_{s}^{2} \left[  \left( 6B^{2}%
+12B-\pi^{2}\right)  \frac{e^{-B}}{\sinh B}-\right. \\
& \left.  -12\ln\left(  1-e^{-2B}\right)  +\frac{6}{\tanh B}\,\mathrm{Li}%
_{2}\left(  e^{-2B}\right)  \right]  ~.
\end{align*}
In particular, we see that at $B=0$ we have $2\operatorname{Im}\Delta( S=1,B=0
) \simeq6.6\,\alpha_{s}^{2}$, which for a typical value of $\alpha_{s}$ is
well below saturation. At $B=0$ the saturation line starts for $\bar{\alpha
}_{s}\ln S\simeq1$, as can be seen from the integral expression (\ref{eq400})
for the phase. The asymptotic linear regime (\ref{eq1200}) is reached for $\ln
S\gtrsim2/\bar{\alpha}_{s}$, with $i\nu_{\mathrm{s}}\simeq0.26$ and
\[
\omega\simeq2.44~\bar{\alpha}_{s}~.
\]
The above are clearly leading order results. However, as we shall explain in more
detail in the next section, the experimental value of $\omega$ in
DIS experiments is lower. For example, in the analysis of  \cite{geometric2} one finds  $\omega \simeq 0.14$, 
since the scaling variable $\tau$ has the form (\ref{eq8100}) with 
$2\omega/\left( 1-\omega\right)  \simeq0.32$. Therefore, as is well known, next to leading
order corrections to the leading BFKL results (which also distinguish between
QCD and its supersymmetric extensions) are important to match to experiment.

\section{Deep Inelastic Scattering in QCD at Small $x$}

We now explore the phenomenological consequences of our results on deeply
saturated cross sections for DIS in QCD at small Bjorken $x$. Throughout the
discussion, we shall assume that we are working in the conformal setting, thus
neglecting the running of the coupling constant and all quark masses. We will associate the scalar
operators $\mathcal{O}_{1}$ and $\mathcal{O}_{2}$ respectively to the photon
and proton. Note that, deep into saturation, the spin of the external
particles plays a minor role, since amplitudes are dominated by the black disk
region with $\sigma( s,\rho,\bar{\rho},\mathbf{b})\simeq1$. As usual, $Q^{2}$
is the photon virtuality and $s\simeq Q^{2}/x$. Moreover, the scale $\bar{Q}$
will now represent a phenomenological parameter, related to the proton
wavefunction, of the order of the relevant proton scales. The wave functions
$f_{1},f_{3}$ and $f_{2},f_{4}$ are localized respectively around $\rho\sim
Q^{-1}$ and $\bar{\rho}\sim\bar{Q}^{-1}$. Therefore, the total cross section
$\Sigma( s,Q,\bar{Q}) $ in (\ref{eq600}) can be approximately computed using
the saturated cross section $\sigma( s,\rho,\bar{\rho}) $ in (\ref{eq5100})
whenever
\begin{equation}
\left\vert \ln\big( Q/\bar{Q}\big) \right\vert \lesssim\mathrm{B}_{s}%
(s/Q\bar{Q})~.\label{eq9000}%
\end{equation}
Moreover, for large $s/Q\bar{Q}$, the saturation line $\mathrm{B}_{s}(S)$ is
approximately linear and $\sigma(s,\rho,\bar{\rho})$ is given by the simple
expression (\ref{eq2000}). In this case, we may easily compute the radial
integrals in (\ref{eq600}) since, on purely dimensional grounds, we must have
that
\[
\int\frac{d\rho}{\rho^{3}}\,f_{1}(\rho)\,f_{3}(\rho) \;\rho^{\zeta}
=Q^{-\zeta}~\gamma(\zeta)
\]
for some constants $\gamma(\zeta)$ of order unity, and similarely for the
proton wave functions. Hence, in the deeply saturated regime at high
$s/Q\bar{Q}$, we expect a rather simple form for the total cross section
$\Sigma(s,Q,\bar{Q})$. Recalling that the cross section $\Sigma$ is
proportional to $Q^{-2}F_{2}$, where $F_{2}\big(x,Q^{2}\big)$ is the usual DIS
proton structure function, we obtain that
\begin{equation}
F_{2}\big( x,Q^{2}\big) \simeq c~\frac{Q}{\Lambda}\left[  \left(  \frac
{Q}{x\Lambda}\right)  ^{\omega}+\left(  \frac{Q}{x\Lambda}\right)  ^{-\omega
}\right]  -\tilde{c}~\frac{Q}{\ \tilde{\Lambda}}\left[  \frac{Q}%
{\tilde{\Lambda}}+\frac{\tilde{\Lambda}}{Q}\right]  ~,\label{magic}%
\end{equation}
where the constants $c,\tilde{c}$ and the scales $\Lambda,\tilde{\Lambda}$ are
the only remenants of our lack of precise knowledge of the scattering radial
wave functions. In particular, $\Lambda,\tilde{\Lambda}$ will be of the same
order as $\bar{Q}$. Had we included the spin of the particles in the
discussion, the parameters $c,\tilde{c},\Lambda,\tilde{\Lambda}$ would carry
also this kinematical information. The exponent $\omega$ is, on the other
hand, universal and depends uniquely on the spin of the pomeron. Note that,
since $\omega\ll1$, the constants of order unity coming from the first two
terms of (\ref{eq2000}) are essentially identical, and we may safely take
\[
\Lambda\simeq\bar{Q}
\]
in first approximation.

As in (\ref{eq2000}), when $\mathrm{B}_{s}(s/Q\bar{Q})\gg\left\vert \ln
(Q/\bar{Q})\right\vert \gg1$, the cross section $\Sigma$ is dominated by the
first term in (\ref{magic}) and exhibits the
geometric scaling
\begin{equation}
\Sigma\sim\frac{1}{\bar{Q}^{2}}~\tau^{-\frac{1-\omega}{2}}~,\label{eq8300}%
\end{equation}
where we define the scaling variable $\tau$ as usual as \cite{geometric}
\[
\tau=\frac{Q^{2}}{Q_{\mathrm{s}}^{2}}%
,~\ \ \ \ \ \ \ \ \ \ \ \ \ \ Q_{\mathrm{s}}^{2}=\bar{Q}^{2}\left(  \frac
{1}{x}\right)  ^{\frac{2\omega}{1-\omega}}~.
\]
Recall that the power of $1/x$ in the saturation scale $Q_{\mathrm{s}}^{2}$ is
observed experimentally, following  \cite{geometric2}, to be given by 
 $2\omega/\left(  1-\omega\right) =  0.321\pm0.056$, so that
\begin{equation}
\omega = 0.138\pm0.021~.\label{eq9100}%
\end{equation}
Deeply into saturation, we predict that $\Sigma$ evolves with $\tau$ with the
specific exponent in (\ref{eq8300}), which is \textit{uniquely} fixed by the
measurement of $\omega$ at the saturation scale $Q_{\mathrm{s}}^{2}$. This has
to be contrasted with the behavior of $\Sigma$ near saturation following from
(\ref{eq8000}), where the exponent of $\tau$ is \textit{not} fixed uniquely by
$\omega$.

\begin{figure}
\begin{center}
\includegraphics[height=9cm]{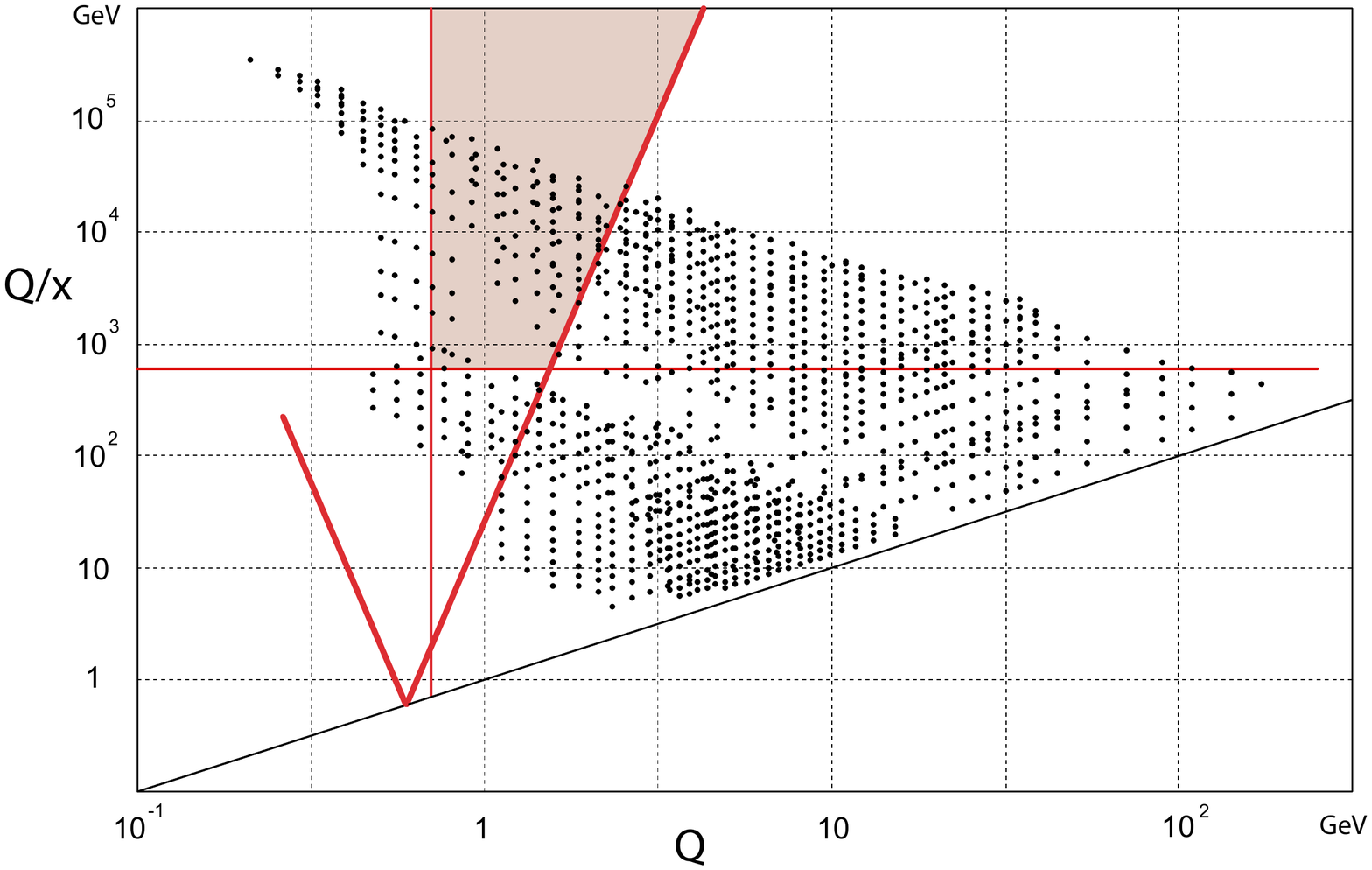}
\end{center}
\caption{
Shown are the available measurements of $F_2(x,Q^2)$ in \cite{D1}, in the $\log_{10}(Q)$--$\log_{10}(Q/x)$ plane,
with energies measured in GeV. All points
lie above the line $x=1$. Shown is also the shaded region of points considered when analyzing (\ref{magic}). It corresponds to the region delimited by the vertical line, setting $Q>Q_{\mathrm{min}}$, the horizontal one, setting $\log_{10}(Q/x\bar{Q})>\eta$, and the asymptotic linear saturation line. This is shown with a thicker line and is
obtained by offsetting the graph in figure \ref{lnS-B} by $\log_{10}(\bar{Q})$ along the $x=1$ line.
}
\label{figDataPoints}
\end{figure}

We wish to test this prediction against the available experimental data on
$F_{2}\big( x,Q^{2}\big)$. Measurements have been performed at values of $x$
and $Q^{2}$ shown in Figure \ref{figDataPoints}, as discussed in \cite{NNPDF},
which collects all available data from \cite{D1}. In this Figure, we present
the data in the $\ln Q$--$\ln Q/x$ plane, with all energy scales measured in
$\mathrm{GeV}$ from now on. These are the natural variables to discuss
saturation, since they enter directly into (\ref{eq9000}). We will fit the
avaliable $F_{2}$ values using (\ref{magic}) in its plausible region of
validity. First of all, we will take
\[
Q>Q_{\text{min}}~
\]
with $Q_{\text{min}}\sim1$ $\mathrm{GeV}$ so that, in first approximation, the
running of the coupling can be neglected. Secondly, we wish to choose points
inside the saturation line (\ref{eq9000}). The exact determination of this
line depends crucially on the phenomenological parameter $\bar{Q}$, and in
turn on the strongly coupled dynamics of the proton. We expect the value of
$\bar{Q}$ to be in the range of available scales -- i.e. the QCD scale and the
proton mass. Assuming that radial wave functions are localized around
$\rho\sim Q^{-1}$ and $\bar{\rho}\sim\bar{Q}^{-1}$, the saturation line in the
$\ln Q$--$\ln Q/x$ plane is then given by the saturation line for the AdS
phase $\Delta$ shown in Figure \ref{lnS-B}, where we replace $B$ and $\ln S$
respectively by $\left\vert \ln Q/\bar{Q}\right\vert $ and $\ln Q/( x\bar{Q}%
)$. In practice, this ammounts to drawing the saturation line of Figure
\ref{lnS-B} onto Figure \ref{figDataPoints}, offsetting the origin along the
line $x=1$ by $\ln\bar{Q}$. We shall then take points with
\[
\omega\, \ln\frac{Q}{x\bar{Q}} >\ln\frac{Q}{\bar{Q}}%
~,~\ \ \ \ \ \ \ \ \ \left(  \bar{Q}\sim0.2-1~\mathrm{GeV}\right)  \ .
\]
Thirdly, we wish to consider data points with high values of $Q/(\bar{Q}x)$,
so as to be into the linear regime of the saturation line. As explained in the
previous section, the leading order BFKL\ analysis suggests that the linear
regime starts around $\ln S\gtrsim2/\bar{\alpha}_{s}$, so that we shall take
data with
\[
\frac{Q}{\bar{Q}x} \gtrsim10^{\eta}~,~\ \ \ \ \ \ \left(  \eta\gtrsim3\right)
\ .
\]

To proceed, let us choose $Q_{\text{min}}=0.7$ $\mathrm{GeV}$, $\bar
{Q}=0.6~\mathrm{GeV}$ and $\eta=3$. We will show later that the main results
are insensitive to this specific choice. The selected data is shown in the
shaded region of Figure \ref{figDataPoints}. We shall test our theoretical
prediction against the real data in \cite{D1} as well as against the very
accurate neural network interpolation to world DIS data in \cite{NNPDFData}. In
particular, we shall minimize the average square deviation of the data from the predicted
theoretical form (\ref{magic}).  Since the parameters $c$, $\tilde{c}$ and $\tilde{c}/\tilde{\Lambda}^2$
enter linearly in (\ref{magic}), minimization reduces to a linear system
parameterized by the single parameter $\omega$. 

\begin{figure}
\begin{center}
\includegraphics[height=5.5cm]{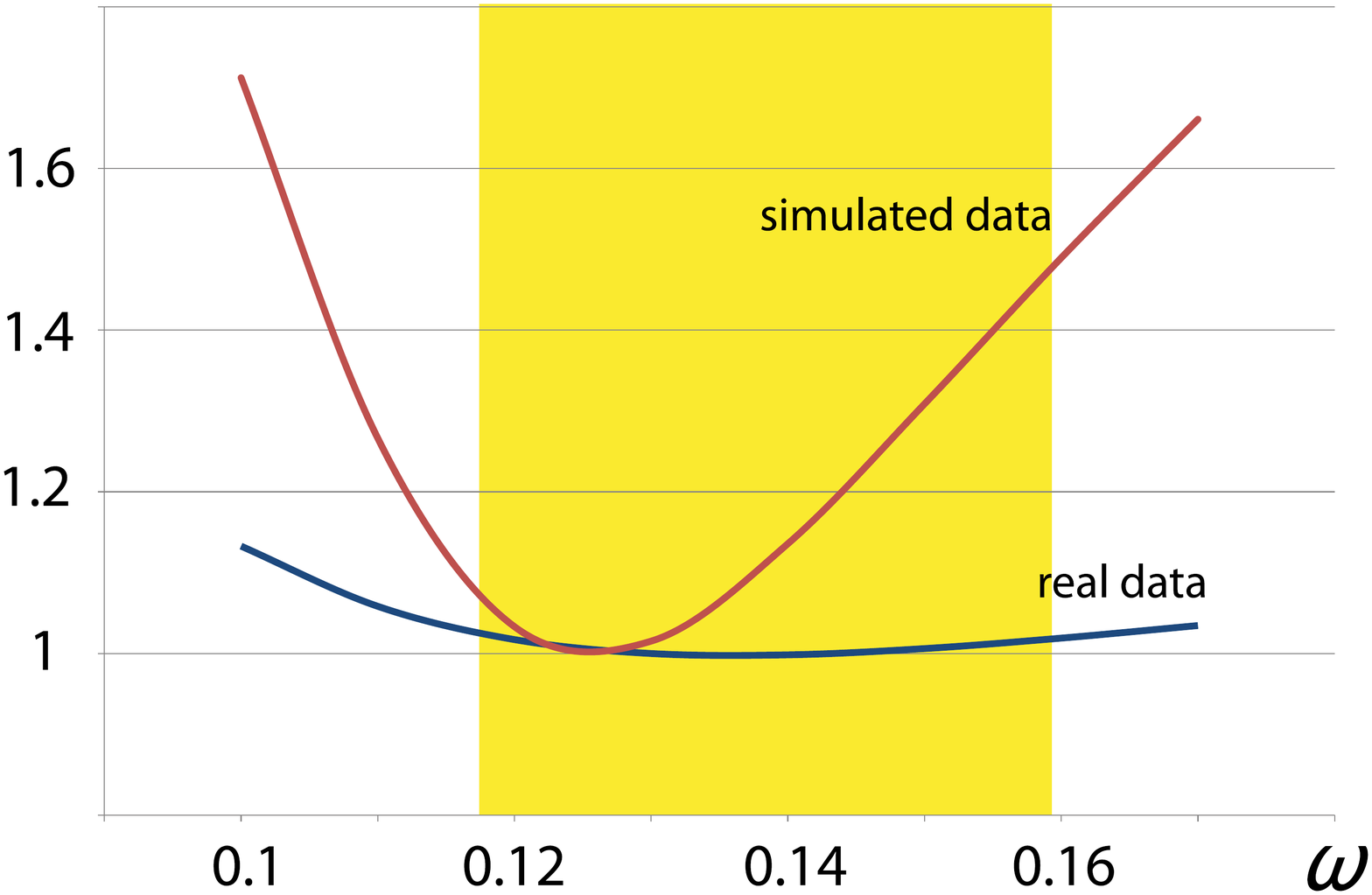}
\end{center}
\caption{Error functions for real and simulated data, plotted as a function of $\omega$. For 
comparison, both functions have been normalized to $1$ at their respective minima. The
error for the simulated data exhibits a sharp minimum, while the corresponding error for real data is more
insensitive to $\omega$, due to a large experimental uncertainty. 
We therefore choose to fix $\omega$ using the simulated data. The yellow stripe
corresponds to the range $\omega = 0.138\pm0.021$
obtained from geometric scaling.}
\label{figError} 
\end{figure}

As a function of $\omega$, the parameters $c$, $\tilde{c}$ and $\tilde{c}/\tilde{\Lambda}^2$ are easily
determined both for real as well as for simulated data. As is clear from
(\ref{magic}), the parameter $\omega$ controls the growth of $F_2$ as $1/x$ increases.
More precisely, at fixed $Q$, the coefficients $c$ and $\omega$
determine the slope as well as the convexity of the function $F_2$ in the 
experimental region of interest, shaded in Figure \ref{figDataPoints}. Unfortunately,
the relevant kinematics is on the boundary of the currently
accessible experimental settings, resulting in data of relatively poor
quality with large experimental uncertainty. This is reflected in
the fact that the error function, although it presents a minimum for $\omega \simeq 0.136$,
is essentially constant in the range $0.1$ to $0.17$ plotted in Figure  \ref{figError}.
On the other hand, as shown in the same figure, if we use the more accurate simulated 
function $F_2$ computed at the same values of  $x$ and $Q^2$ available in the
real data \cite{NNPDFData}, we obtain a rather sharp minimum for the error function at
\[
\omega \simeq 0.126\ .
\]
Therefore, from now on, we shall determine the optimal value of $\omega$ using
the simulated data only. At this point we wish to emphasize that this
value of $\omega$, obtained from data inside the saturated kinematical region, 
is within the experimental range  $\omega = 0.138\pm0.021$,
obtained independently from geometric scaling.
The values for the other relevant parameters can be
determined to be $\tilde{\Lambda}\simeq 1.0~\mathrm{GeV}$, $c\simeq0.13$, 
$\tilde{c}\simeq0.14$ for the real data and $\tilde{\Lambda}\simeq1.0~\mathrm{GeV}$, $c\simeq0.11$, 
$\tilde{c}\simeq0.08$ for the simulated one.

\begin{figure}
\begin{center}
\includegraphics[height=15cm]{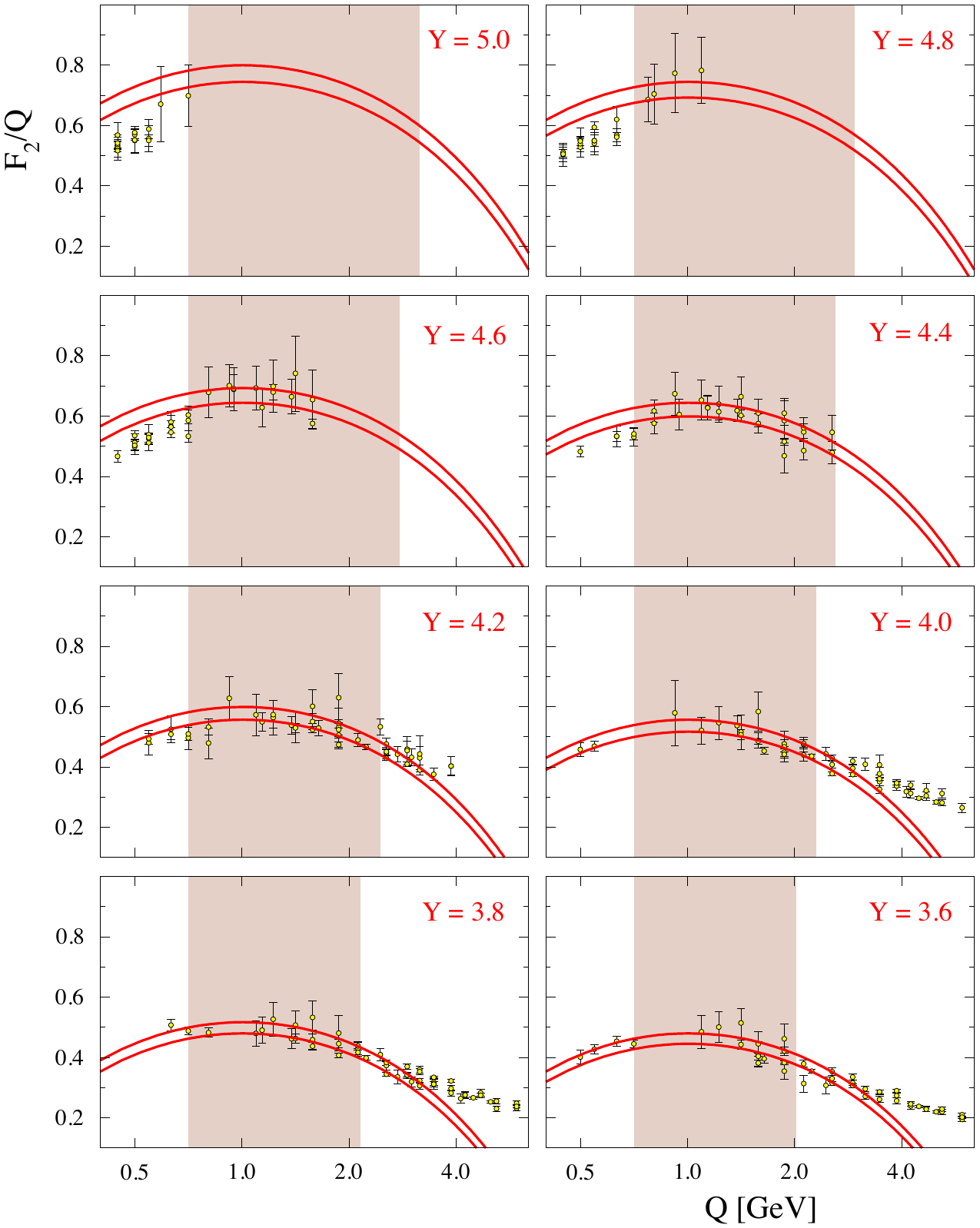}
\end{center}
\caption{Real data  $F_{2}/Q$ as a function of $Q$. Each graph contains data
points with values of $\log_{10}( Q/x)$ in the range $Y\pm0.1$, for 
$3.6 \leq Y \leq 5$ in increments of $0.2$.  Theoretical 
curves are shown in red both for the minimal and maximal values of $\log_{10}( Q/x)$. 
The shaded area corresponds to the region delimited by the choice of parameters
$Q_{\text{min}}=0.7$ $\mathrm{GeV}$, $\bar
{Q}=0.6~\mathrm{GeV}$ and $\eta=3$, as also  shown in Figure \ref{figDataPoints}.}
\label{fit1}
\end{figure}

\begin{figure}
\begin{center}
\includegraphics[height=15cm]{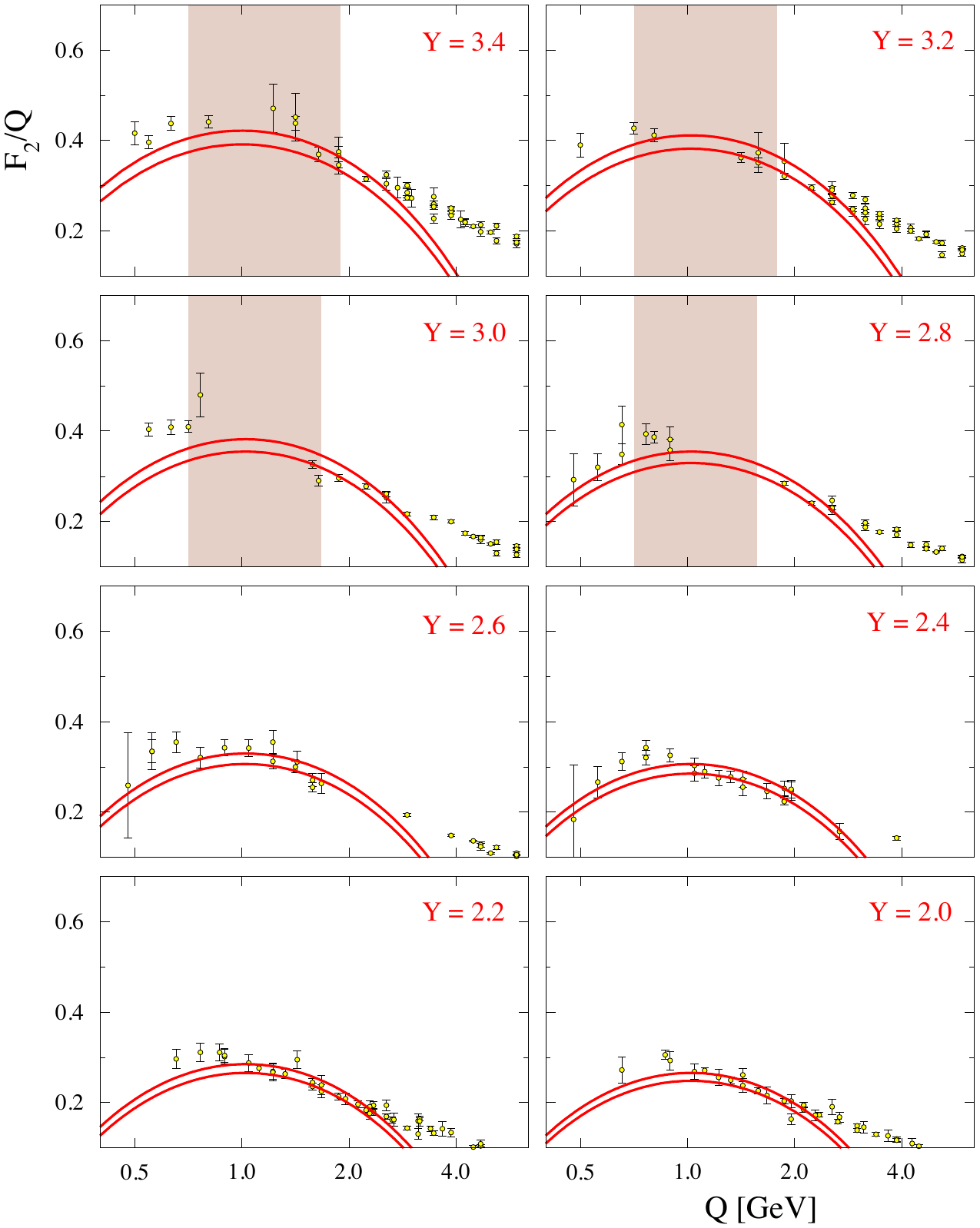}
\end{center}
\caption{Real data  $F_{2}/Q$ as a function of $Q$. 
Each graph contains data points with values of $\log_{10}( Q/x)$ in the range $Y\pm0.1$, for 
$2\leq Y\leq 3.4$ in increments of $0.2$.  Theoretical 
curves are shown in red both for the minimal and maximal values of $\log_{10}( Q/x)$. 
The shaded area corresponds to the region delimited by the choice of parameters
$Q_{\text{min}}=0.7$ $\mathrm{GeV}$, $\bar
{Q}=0.6~\mathrm{GeV}$ and $\eta=3$,  as also  shown in Figure \ref{figDataPoints}.}
\label{fit2}
\end{figure}

\begin{figure}
\begin{center}
\includegraphics[height=15cm]{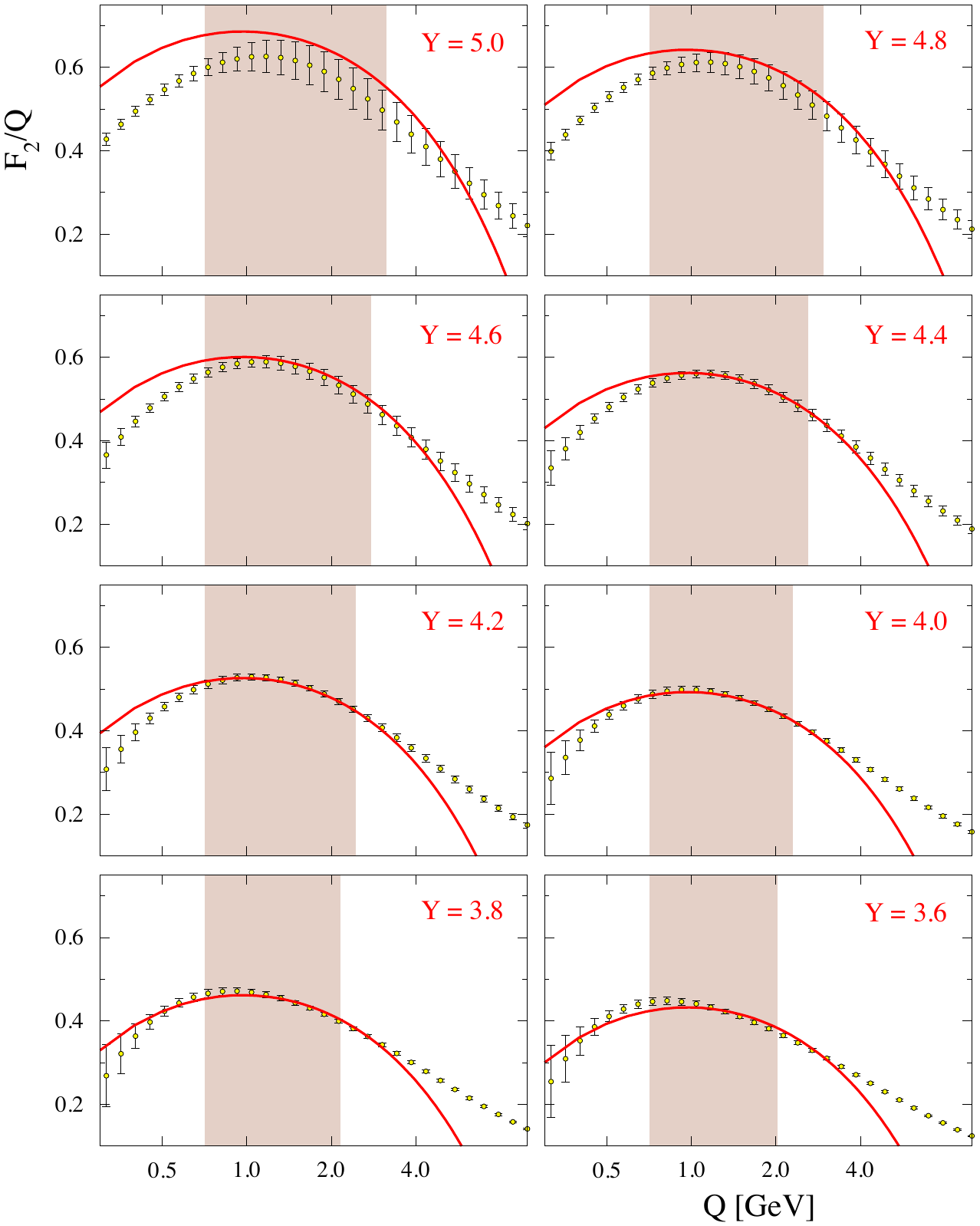}
\end{center}
\caption{Same as Figure \ref{fit1} for the simulated function $F_{2}$ computed at
evenly spaced values of $Q$ for fixed  $Y=\log_{10}( Q/x)$.}
\label{fit3}
\end{figure}

\begin{figure}
\begin{center}
\includegraphics[height=15cm]{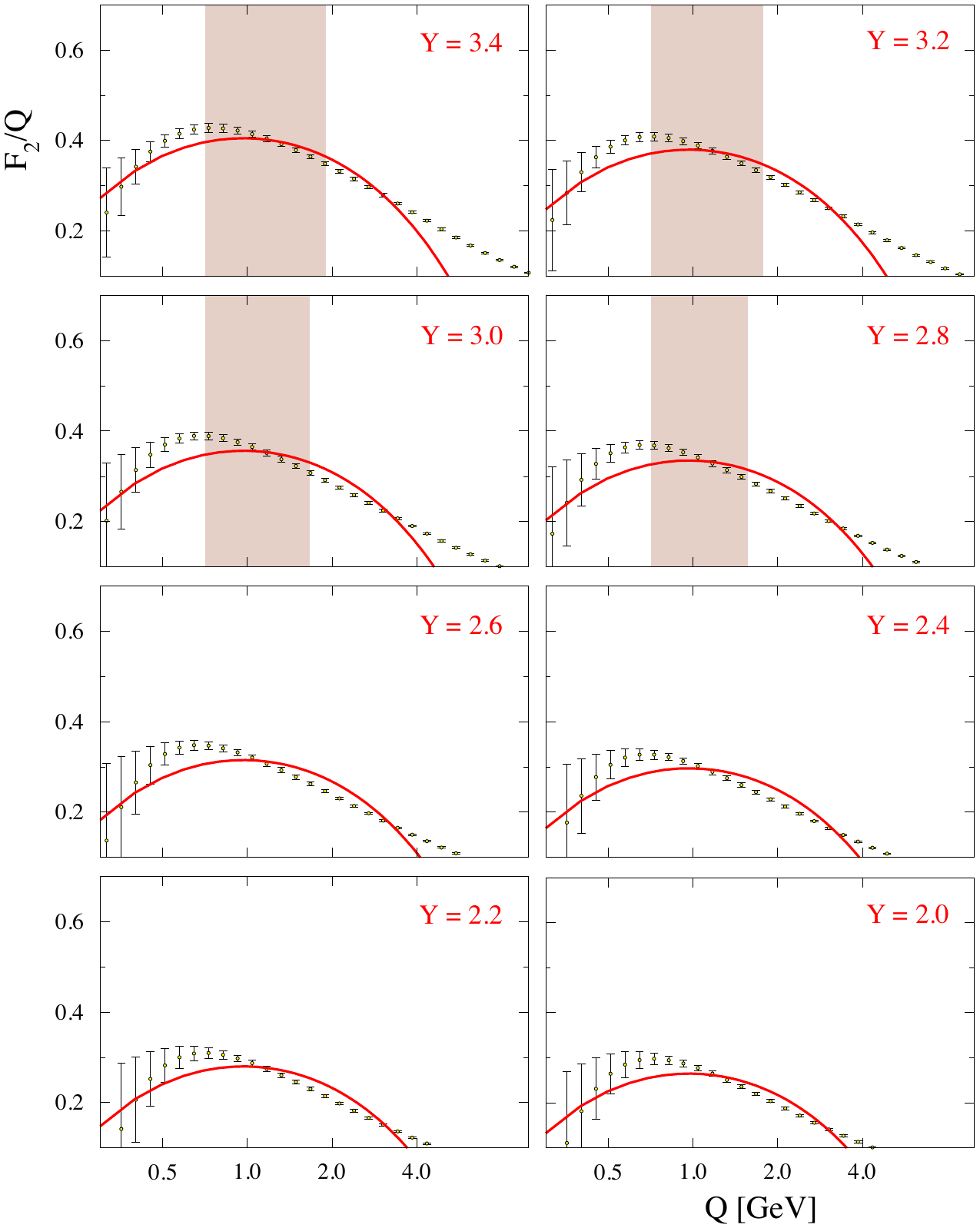}
\end{center}
\caption{Same as Figure \ref{fit2} for the simulated function $F_{2}$ computed at
evenly spaced values of $Q$ for fixed  $Y=\log_{10}( Q/x)$.}
\label{fit4}
\end{figure}

The real data is presented in Figures \ref{fit1} and \ref{fit2}, where we
show the data together with the theoretical curves from (\ref{magic}). We
plot $F_{2}/Q$ as a function of $\log_{10}Q$, and each graph contains data
points with values of $\log_{10}( Q/x)$ in the range $Y\pm0.1$, for $2\leq
Y\leq5$ in increments of $0.2$. Theoretical 
curves are shown in red both for the minimal and maximal values of $\log_{10}( Q/x)$. 
Finally, the shaded area corresponds to the region delimited by the choice of parameters
$\bar{Q},\eta$, $Q_{\text{min}}$, as also shown in Figure \ref{figDataPoints}. Analogously,
Figures \ref{fit3} and \ref{fit4} show the simulated data.

Let us note that the AdS black disk form of the structure function given in
(\ref{magic}), with the above choice of parameters, approximates the available real
data with an average $6\%$ accuracy in the rather large region of parameter
space $0.5<Q^{2}<10$ and $x<10^{-2}$.

\begin{table}[!h]
\caption{Number of experimentally available data points $n$ and predicted value of $\omega$ for different values of $\bar{Q}$ and $Q_{\text{min}}$.}
\label{tb}
\begin{center}
\begin{tabular}[!t]{|c|c|c||c|c|}
\hline
$\  \  \bar{Q}\ \ $ & $ Q_{\text{min}} $ & $^{\phantom{\frac{1}{1}}}\eta^{\phantom{\frac{1}{1}}}$ &$n$ & $\omega$\\
\hline\hline
0.3 & 0.7 & 3 &58 &  0.104\\
\hline
0.3 &  1 & 3 &23 & 0.090\\
\hline
0.6 &  0.7 & 3 &138 & 0.126\\
\hline
0.6 &  1 & 3 &104 & 0.130\\
\hline
1 &  0.7 & 3 &200 &0.141\\
\hline
1 &  1 & 3 &171 & 0.152\\
\hline
\end{tabular}
\end{center}
\end{table}

Due to uncertainty on the precise location of the saturation line, we have
repeted the analysis with different values of $\bar{Q}$, $Q_{\text{min}}$ and
$\eta$, to test the robustness of the predicted value for $\omega$. Within the
range $0.7<Q_{\text{min}}<1$ and $0.3<\bar{Q}<1$, the fitted value for
$\omega$ varies from $0.090$ to $0.152$, as shown in Table \ref{tb}, thus
mostly within the predicted range (the optimal value of $\omega$ is rather
insensitive to the choiche of $\eta>3$ which we keep fixed). Note that, although the
first two entries of Table \ref{tb} are outside the predicted range, they are based on a very small number
of data points.

As already stressed, avaliable data is on the boundary of the deeply saturated region, and
one would need to reach higher energies in order to better test these
predictions. Possibly, future data from LHC will be of use to confirm the
above results.

\section{Relation to the Dipole Formalism}

We will conclude this paper by discussing the relation between the above
results and the dipole formalism \cite{dipole}, which is usually employed in
the analysis of saturation effects. In this context, it is costumary to
analyze the so--called dipole--dipole cross section $\sigma_{\mathrm{DD}}(
s,\mathbf{r},\mathbf{\bar{r}},\mathbf{b})$ instead of $\sigma( s,\rho
,\bar{\rho},\mathbf{b})$, with the full cross section $\Sigma( s,Q,\bar{Q})$
given by integrals over the dipole transverse orientations $\mathbf{r}$ and
$\mathbf{\bar{r}}$
\begin{align}
& \frac{2}{\left(  2\pi\right)  ^{2}}\int\frac{d^{2}\mathbf{r}}{\mathbf{r}%
^{4}} \mathbf{~}\frac{d^{2}\mathbf{\bar{r}}}{\mathbf{\bar{r}}^{4}}\;\; W(
\mathbf{r}) \;\; \sigma_{\mathrm{DD}}( s,\mathbf{r},\mathbf{\bar{r}})\;\;
\bar{W}( \mathbf{\bar{r}}) ~,\label{eq6000}\\
& \sigma_{\mathrm{DD}}( s,\mathbf{r},\mathbf{\bar{r}}) = \int d^{2}%
\mathbf{b}\;\;\sigma_{\mathrm{DD}}( s,\mathbf{r},\mathbf{\bar{r}},\mathbf{b})
~,\nonumber
\end{align}
where $W(\mathbf{r})$ and $\bar{W}( \mathbf{\bar{r}})$ are the so-called dipole impact factors.

Let us first note that, although the dipole formalism is quite useful due to its intuitive
physical description of the high energy process and of the linear BFKL and
nonlinear BK evolutions \cite{MuellerLectures}, it is not well suited for the
discussion of unitarization, since the natural object which satisfies the
unitarity constraint $0\leq\sigma\leq2$ is $\sigma(s,\rho,\bar{\rho},\mathbf{b})$, 
instead of $\sigma_{\mathrm{DD}}(s,\mathbf{r},\mathbf{\bar{r}%
},\mathbf{b})$. This fact is quite clear in gauge theories which are exactly conformal,
like $\mathcal{N}=4$ super Yang--Mills, where
the dipole formalism can still be applied (as well as the BK\ equation, which is explicitly conformally invariant).
In this case, the theory has no asymptotic states or an S-matrix to which to
apply the usual unitarity constraints. Moreover,
even in a confining theory like QCD, which possesses asymptotic states, the dipole
state is not a single particle state at infinity, and therefore does not
enter in a usual S-matrix element. In fact, in the standard discussions of DIS at small $x$,
the dipole picture is often used to describe the wave function of an off-shell spacelike photon.

At any rate, in order to make contact with the standard
litterature, we will briefly analyze, in what follows, the above expressions
in the unsaturated regime of small AdS phase shift $\left\vert \Delta\right\vert \ll 1$.

\begin{figure}
\begin{center}
\includegraphics[height=5cm]{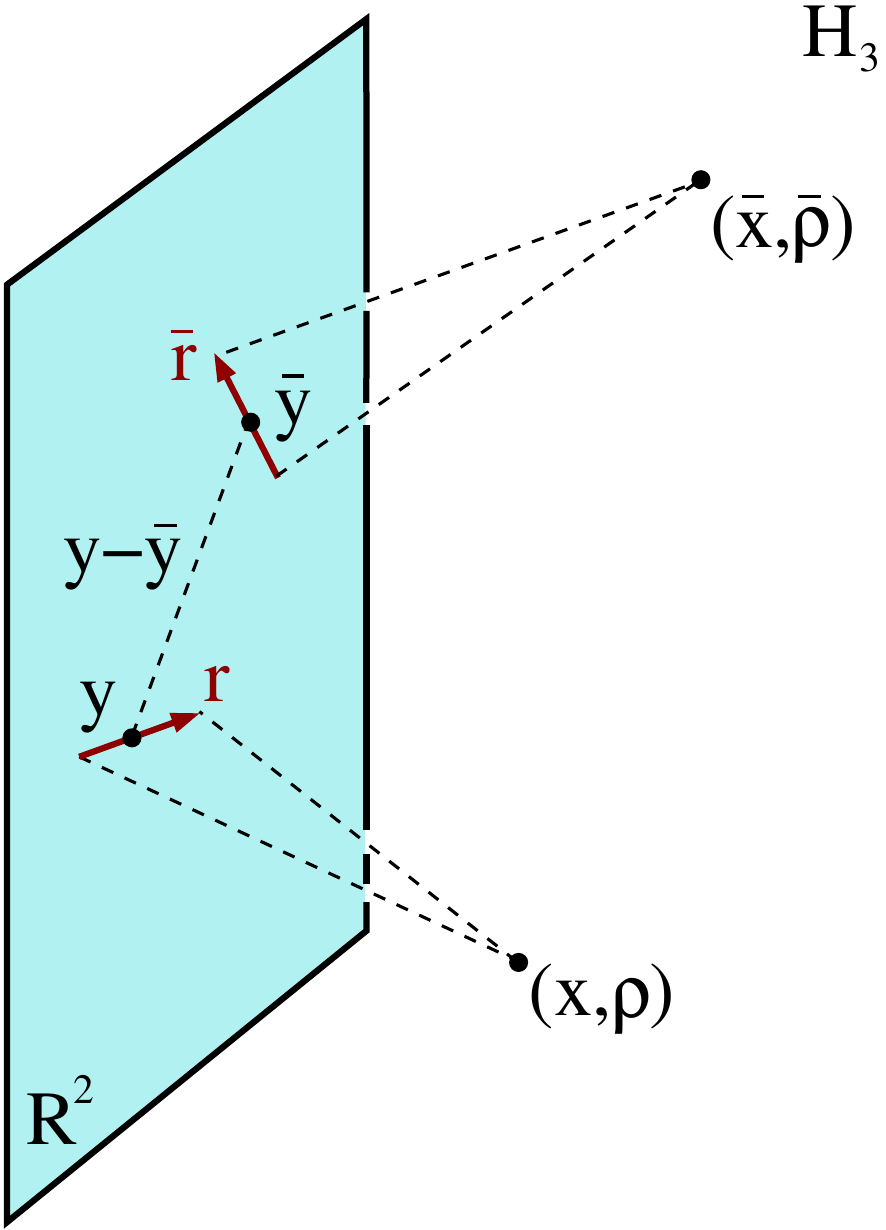}
\end{center}
\caption{Relation to dipole formalism. While the dipole--dipole cross section 
$\sigma_{\mathrm{DD}}(s,\mathbf{r},\mathbf{\bar{r}},\mathbf{y}-\bar{\mathbf{y}})$ 
depends on four points in \textrm{R}$^2$, the cross section 
$\sigma( s,\rho,\bar{\rho},\mathbf{x}-\bar{\mathbf{x}})$,
which is the one constrained by unitarity, depends only on two
points in \textrm{H}$_{3}$. The role of the dipole vectors
$\mathbf{r}$ and $\mathbf{\bar{r}}$ is now played by the radial coordinates
$\rho$ and $\bar{\rho}$.}
\label{dipole}
\end{figure}

Let us first analyze the impact factors $W(\mathbf{r})$ and $\bar{W}(\mathbf{\bar r})$, 
leaving to the second part of this section the discussion on $\sigma_{\mathrm{DD}}(
s,\mathbf{r},\mathbf{\bar{r}},\mathbf{b})$ and on saturation in the context of the dipole formalism.
We recall the BFKL representation of $\Delta$ analyzed in \cite{Paper5}. More
precisely, to leading order in the coupling we have that
\[
2\Delta=\frac{i}{N^{2}}\int\frac{d^{2}\mathbf{r}}{\mathbf{r}^{4}}\,
\frac{d^{2}\mathbf{\bar{r}}}{\mathbf{\bar{r}}^{4}}\, d^{2}\mathbf{y~}%
d^{2}\mathbf{\bar{y}~}\; W(\rho,\mathbf{r},\mathbf{x}-\mathbf{y}) \;\; 
F(\mathbf{r,\bar{r},y-\bar{y}})\;\; \bar{W}( \bar{\rho},\mathbf{\bar{r},\bar{x}-\bar
{y}}) ~,
\]
with $\mathbf{b}=\mathbf{x}-\mathbf{\bar{x}}$. The impact factor $W$ depends
on the point $\rho,\mathbf{x}$ in \textrm{H}$_{3}$ and on the two intermediate
points $\mathbf{y\pm r}/2$ on the boundary of \textrm{H}$_{3}$, as represented in 
Figure \ref{dipole}. Similar
comments apply to the impact factor $\bar{W}$. Moreover, 
$F(\mathbf{r,\bar{r},y-\bar{y}}) $ is the leading order two--gluon exchange kernel from
$\mathbf{y}\pm\mathbf{r}/2$ to $\mathbf{\bar{y}}\pm\mathbf{\bar{r}}/2$.
Integrating against
\[
2\int d^{2}\mathbf{b} \int\frac{d\rho}{\rho^{3}}\,f_{1}( \rho)\,f_{3}( \rho)
\int\frac{d\bar{\rho}}{\bar{\rho}^{3}}\,f_{2}(\bar{\rho})\, f_{4}\left(
\bar{\rho}\right)
\]
and using the approximate relation (\ref{eq4000}) valid in the small phase
regime, we obtain an expression of the form (\ref{eq6000}), where
\begin{align}
\sigma_{\mathrm{DD}}( s,\mathbf{r},\mathbf{\bar{r}})  & \simeq c\bar{c}%
\,\frac{\left(  2\pi\right)  ^{2}}{N^{2}}\, \operatorname{Re}\int
d^{2}\mathbf{w}\; F( \mathbf{r,\bar{r},w}) ~,\label{eq7100}\\
W( \mathbf{r})  & =\frac{1}{c} \int\frac{d\rho}{\rho^{3}}\,f_{1}( \rho)\,
f_{3}( \rho) \int d^{2}\mathbf{w}\; W( \rho,\mathbf{r}, \mathbf{w})
~,\nonumber
\end{align}
and similarely for $\bar{W}( \mathbf{\bar{r}}) $. The constants $c,\bar{c}$
are fixed by the normalization conditions
\[
\frac{1}{2\pi}\int\frac{d^{2}\mathbf{r}}{\mathbf{r}^{4}}\;W( \mathbf{r})
=\frac{1}{2\pi}\int\frac{d^{2}\mathbf{\bar{r}}}{\mathbf{\bar{r}}^{4}}\;
\bar{W}( \mathbf{\bar{r}}) =1
\]
analogous to (\ref{eq7000}).

Let us discuss the impact factor $W$ in detail. As shown in \cite{Paper5},
conformal invariance highly constrains $W(\rho, \mathbf{r},\mathbf{w})$ to be
a function of the unique cross ratio
\[
\frac{\mathbf{r}^{2}\rho^{2}}{\left[  \rho^{2}+\left(  \mathbf{w-}%
\frac{\mathbf{r}}{2}\right)  ^{2}\right]  \left[  \rho^{2}+\left(
\mathbf{w+}\frac{\mathbf{r}}{2}\right)  ^{2}\right]  }~,
\]
which can be conveniently written in the following integral representation\footnote{In the 
notation of \cite{Paper5}, $W( \nu) $ is given by $V( \nu)
/V_{\text{min}}( \nu,1)$, with $V( \nu) $ the impact factor for the full
amplitude and with $V_{\text{min}}(\nu,1) =\Gamma\left(  \frac{2\Delta
_{1}-1+i\nu}{2}\right)  \Gamma\left(  \frac{2\Delta_{1}-1-i\nu}{2}\right)  /
\big( \Gamma(\Delta_{1}) \Gamma(\Delta_{1}-1)\big)$, where $\Delta_{1}$ is the
dimension of the external operator $\mathcal{O}_{1}$. Similarely for $\bar{W}%
$.}
\begin{align}
& \frac{1}{64\pi^{5}}\int d\nu~\nu^{2}\left(  1+\nu^{2}\right)  \frac
{\Gamma^{2}\left(  \frac{1-i\nu}{2}\right)  }{\Gamma\left(  1-i\nu\right)  }
~W( \nu) \times\label{eq6100}\\
& \times\int d^{2}\mathbf{z}\left(  \frac{\rho}{\rho^{2}+\left(
\mathbf{w}-\mathbf{z}\right)  ^{2}}\right)  ^{1+i\nu}\left(  \frac
{\mathbf{r}^{2}}{\left(  \mathbf{z-}\frac{\mathbf{r}}{2}\right)  ^{2}\left(
\mathbf{z+}\frac{\mathbf{r}}{2}\right)  ^{2}}\right)  ^{\frac{1-i\nu}{2}%
}~,\nonumber
\end{align}
where the transforms $W( \nu) $ and $\bar{W}( \nu) $ determine the Regge
residue $\beta( \nu) $ to be
\[
\beta( \nu) =\frac{i}{4\nu}\,W( \nu) \;\tanh(\pi\nu/2)\; \bar{W}( \nu) ~.
\]
Moving to momentum space in the transverse $\mathbb{E}^{2}$ plane by
integrating against $\int d^{2}\mathbf{w}$ we obtain
\begin{equation}
\frac{1}{32\pi^{3}}\int d\nu~\left(  1+\nu^{2}\right)  ~W( \nu) \int_{0}%
^{1}\frac{d\zeta}{\sqrt{\zeta\left(  1-\zeta\right)  }}~~\rho\left\vert
\mathbf{r}\right\vert \left(  \frac{\zeta\left(  1-\zeta\right)
\mathbf{r}^{2}}{\rho^{2}}\right)  ^{\frac{i\nu}{2}}~,\label{eq6200}%
\end{equation}
where $\zeta$ is the Feynman parameter related to the denominators in the last
parenthesis of (\ref{eq6100}).

For concreteness, let us return to the specific example already discussed in
section \ref{secDeep}, with $\beta( \nu) $ given by (\ref{eq10}). We have that
$W( \nu) = 8\pi^{2}\bar{\alpha}_{s}/\left(  1+\nu^{2}\right) $. Then $W(
\mathbf{r} )$ can be computed from (\ref{eq6200}), since the $\nu$ integral
fixes $\rho=\left\vert \mathbf{r}\right\vert \sqrt{\zeta\left(  1-\zeta
\right)  }$. After integrating against $c^{-1} \int d\rho~\rho^{-3}f_{1}f_{3}%
$, we obtain
\[
W( \mathbf{r} ) = \int_{0}^{1}\frac{d\zeta}{\zeta\left(  1-\zeta\right)
}~f_{1}\left(  \sqrt{\zeta\left(  1-\zeta\right)  }\left\vert \mathbf{r}%
\right\vert \right)  f_{3}\left(  \sqrt{\zeta\left(  1-\zeta\right)
}\left\vert \mathbf{r}\right\vert \right)  ~,
\]
with $c=\bar{\alpha}_{s}/2$. Using the fact that $f_{i}( \rho) =\sqrt{2}%
Q_{i}\rho^{2}K_{0}( Q_{i}\rho) $ for $i=1,3$, we have just obtained the usual
expression in terms of dipole wave functions \cite{dipole}.

To conclude this section, let us recall the form of the dipole--dipole cross section
$\sigma_{\mathrm{DD}}( s,\mathbf{r},\mathbf{\bar{r}})$, in the unsaturated
case (\ref{eq7100}). It is given by the usual BFKL kernel%
\[
\sigma_{\mathrm{DD}}( s,\mathbf{r},\mathbf{\bar{r}}) \simeq\frac{\left\vert
\mathbf{r}\vert\vert\mathbf{\bar{r}}\right\vert }{2\pi N^{2}} \int d\nu
\;\frac{16\pi^{3}\bar{\alpha}_{s}^{2}}{\left(  1+\nu^{2}\right) ^{2}}\; \left(
s\vert\mathbf{r}\vert\vert\mathbf{\bar{r}}\vert\right) ^{j\left(  \nu\right)
-1} \left\vert \frac{\mathbf{\bar{r}}}{\mathbf{r}}\right\vert ^{-i\nu}~~,
\]
which should be compared with (\ref{eq7200}), with $\beta(\nu) $ given by (\ref{eq10}). Recall also that
the above expression is the zero momentum contribution to the full BFKL expression for $\sigma
_{\mathrm{DD}}\left(  s,\mathbf{r},\mathbf{\bar{r}},\mathbf{b}\right)  $ given
by \cite{Lipatov}
\begin{equation}
\sigma_{\mathrm{DD}}\left(  s,\mathbf{r},\mathbf{\bar{r}},\mathbf{b}\right)
\simeq \frac{i}{2\pi^2 N^2}\int d\nu~\nu~\alpha\left(  \nu\right)  ~\left(
s\vert\mathbf{r}\vert\vert\mathbf{\bar{r}}\vert\right) ^{j\left(  \nu\right)
-1}~\mathcal{T}_{i\nu}\left(  \mathbf{r},\mathbf{\bar{r}},\mathbf{b}\right)
~,\label{eq1}%
\end{equation}
where
\[
\alpha\left(  \nu\right) =\frac{16\pi^3\, \bar{\alpha}_s^2}{\left(  1+\nu^{2}\right)  ^{2}}%
\,\frac{\Gamma\left(1-i\nu\right) }{\Gamma\left(  \frac{1-i\nu
}{2}\right)  ^{2}}\,\frac{\Gamma\left(  \frac{1+i\nu}{2}\right)  ^{2}}{\Gamma\left( 1+i\nu\right)  }~,
\]
and where $\mathcal{T}_{i\nu
}\left(  \mathbf{r},\mathbf{\bar{r}},\mathbf{b}\right)  $ is the
$2$--dimensional conformal partial wave of spin $0$ and conformal dimension
$1+i\nu$ at the four points $(\mathbf{b\pm r})/2$, $(-\mathbf{b\pm
\bar{r}})/2$. Due to transverse conformal invariance, $\mathcal{T}_{i\nu}$ depends
uniquely on the cross--ratio combinations%
\begin{align*}
z\bar{z}  &  =\frac{\mathbf{r}^{2}\mathbf{\bar{r}}^{2}}{\left(  \mathbf{b-}%
\frac{\mathbf{r}}{2}+\frac{\mathbf{\bar{r}}}{2}\right)  ^{2}\left(
\mathbf{b+}\frac{\mathbf{r}}{2}-\frac{\mathbf{\bar{r}}}{2}\right)  ^{2}}\ ,\\
\frac{z\bar{z}}{\left(  1-z\right)  \left(  1-\bar{z}\right)  }  &
=\frac{\mathbf{r}^{2}\mathbf{\bar{r}}^{2}}{\left(  \mathbf{b-}\frac
{\mathbf{r}}{2}-\frac{\mathbf{\bar{r}}}{2}\right)  ^{2}\left(  \mathbf{b+}%
\frac{\mathbf{r}}{2}+\frac{\mathbf{\bar{r}}}{2}\right)  ^{2}}\ ,
\end{align*}
and is given explicitly by
\begin{align*}
\mathcal{T}_{i\nu}\left(  \mathbf{r},\mathbf{\bar{r}},\mathbf{b}\right)   &
=\left( -z\right)  ^{h}\left(  -\bar{z}\right)  ^{h}F\left(  h,h,2h,z\right)  F\left(
h,h,2h,\bar{z}\right)  ~,\\
h  &  =\frac{1+i\nu}{2}~,
\end{align*}
where $F$ is the hypergeometric function $\,_2 F_1$. 

The expression (\ref{eq1}) should be confronted 
with $\sigma\left(  s,\rho,\bar{\rho},\mathbf{b}\right)$
derived from  equations (\ref{eq4000}) and (\ref{eq400})
in the limit of small AdS phase shift $\left\vert \Delta\right\vert \ll 1$,
\begin{equation}
\sigma\left(  s,\rho,\bar{\rho},\mathbf{b}\right)  \simeq \frac{i}{2\pi^2 N^2}
\int d\nu
~\nu~\operatorname{Im}\beta\left(  \nu\right)  ~S^{j\left(  \nu\right)  -1}~\frac{e^{-i\nu B}%
}{\sinh B}\ .
\label{c1}
\end{equation}
Given the similarity of the expressions for $\sigma_{\mathrm{DD}}$ and $\sigma$, we may try to follow
the program of section 2 and consider the saturation region in the transverse impact parameter $\mathbf{b}$--space
where, at fixed energy $s$ and dipole orientations $\mathbf{r}, \mathbf{\bar r}$, 
the cross section $\sigma_{\mathrm{DD}}\left(  s,\mathbf{r},\mathbf{\bar{r}},\mathbf{b}\right)$ becomes greater then unity.
On the other hand, this program cannot be carried out in general. In fact, while the cross section $\sigma\left(  s,\rho,\bar{\rho},\mathbf{b}\right)$
depends, aside from energy, on a single conformal cross ratio $B$, the cross section $\sigma_{\mathrm{DD}}\left(  s,\mathbf{r},\mathbf{\bar{r}},\mathbf{b}\right)$ 
depends on energy and on two cross ratios $z,\bar{z}$ which reflect the orientations of the two dipoles
(the scalar quantities $\rho,\bar\rho$ are replaced by the dipole transverse vectors $\mathbf{r},\mathbf{\bar{r}}$). Moreover,
both $B$ and $\operatorname{ln}S$
enter exponentially in (\ref{c1}) and allow for a simple saddle approximation of the integral and 
for the determination of the saturation line as in (\ref{eq1200}). On the other hand, the dependence
of the integrand in (\ref{eq1}) on $z,\bar{z}$ is now highly
non--trivial, since it involves not only the norms but also the orientations of
$\mathbf{r},\mathbf{\bar{r}},\mathbf{b}$. Most importantly, it does not allow
for a simple approximation of the integral at a saddle point and a simple 
determination of the saturation line.

The analysis of the saturation region for $\sigma_{\mathrm{DD}}$ can
be carried out only in the limit $\left\vert \mathbf{r}\right\vert ,\left\vert
\mathbf{\bar{r}}\right\vert \ll\left\vert \mathbf{b}\right\vert $. 
In fact, in this limit, one may use the operator product expansion and obtain
\begin{equation}
\mathcal{T}_{i\nu}\left(  \mathbf{r},\mathbf{\bar{r}},\mathbf{b}\right)
\simeq\left(  \frac{\left\vert \mathbf{r}\right\vert \left\vert \mathbf{\bar
{r}}\right\vert }{\mathbf{b}^{2}}\right)  ^{1+i\nu}\label{eq3}%
\end{equation}
of the simple exponential form. Similarly, in the same limit
$\rho,\bar{\rho}\ll\left\vert \mathbf{b}\right\vert $ one
has $B\simeq\ln\left(  \mathbf{b}^{2}/\rho\bar{\rho}\right)  $ and one may
substitute, in (\ref{c1})
\[
\frac{e^{-i\nu B}}{\sinh B}\simeq\left(  \frac{\rho\bar{\rho}}{\mathbf{b}^{2}%
}\right)  ^{1+i\nu}%
\]
obtaining an expression analogous to (\ref{eq3}). From here on we may follow the 
usual steps reviewed in section 2 to determine 
the saturation radius, which is given in general by $\operatorname{ln}(\mathbf{b}^2
/\vert\mathbf{r}\vert\vert\mathbf{\bar{r}}\vert)\simeq \omega \operatorname{ln}(s\vert\mathbf{r}\vert\vert\mathbf{\bar{r}}\vert)$, 
leading to a black disk cross--section given by 
$\pi \mathbf{b}^2\simeq \pi \vert\mathbf{r}\vert\vert\mathbf{\bar{r}}\vert (s\vert\mathbf{r}\vert\vert\mathbf{\bar{r}}\vert)^\omega$. 

With this procedure we recover an expression analogous to the first term of (\ref{eq2000}) just as easily in the dipole formalism.
On the other hand, as we already pointed out, the usual saddling argument works in (\ref{c1}) for
generic values of $\rho,\bar{\rho},\mathbf{b}$. 
This fact allows to determine the other terms in
equation (\ref{eq2000}), which could not have been deduced in the 
dipole language. Note that the extra three terms in (\ref{eq2000}), and correspondingly in (\ref{magic}), are
crucial in order to have a qualitatively good fit for the relevant $F_{2}$
data at hand. In fact, the expression for $F_{2}/Q$
exhibits a non-trivial dependence on $Q$ at fixed $Q/x$, which is qualitatively
correctly captured by the third and fourth terms in (\ref{magic}) proportional to
\[
-\frac{Q}{\tilde{\Lambda}}-\frac{\tilde{\Lambda}}{Q}~.
\]
These terms give a concave behavior with maximum at $Q\simeq\tilde{\Lambda}$,
which is a clear feature of the $F_{2}$ data, as can be seen from the plots in Figures \ref{fit1}, \ref{fit2}, \ref{fit3} and \ref{fit4}.
A pure term of the form $\left(
Q/x\right)  ^{\omega}$, as could be determined by the above arguments also in
the usual dipole formalism, is clearly insufficient to reproduce the $Q$
dependence at fixed $Q/x$ inside the saturation region.

\section*{Acknowledgments}

We would like to thank Jo\~ao Viana Lopes and
Giulia Galbiati for support with the data analysis, Andrea Banfi for discussions
and Markus Diehl for carefully reading the manuscript. 
LC is funded by the
\textit{Museo Storico della Fisica e Centro Studi e Ricerche "Enrico Fermi"} and 
is partially funded by INFN, by the MIUR--PRIN contract 2005--024045--002 and
by the EU contract MRTN--CT--2004--005104. MC is partially funded by the 
FCT-CERN grant POCI/FP/63904/2005. \emph{Centro de F\'{\i}sica do Porto} is partially 
funded by FCT through the POCI program.

\end{document}